\documentclass[aps,prl,twocolumn,superscriptaddress]{revtex4-1}
\usepackage{graphicx}
\usepackage{mathrsfs}
\usepackage{bm}
\usepackage{amsmath}
\usepackage{color}
\usepackage{dcolumn}
\usepackage{epstopdf}
\usepackage{dsfont}
\usepackage{amssymb}
\usepackage{tabularx}
\usepackage{array,ulem}
\usepackage{float}
\usepackage{colordvi}
\usepackage{verbatim}
\usepackage{hyperref}
\hypersetup{
	colorlinks=true,
	linkcolor=blue,
	filecolor=blue,
	urlcolor=blue,
	citecolor=blue,
}

\begin{document}
\title{Interlayer Coupling Induced Topological Phase Transition to Higher Order
}
\author{Lizhou Liu$^*$}
\affiliation{College of Physics, Hebei Normal University, Shijiazhuang, Hebei 050024, China}
\author{Jiaqi An$^*$}
\affiliation{International Centre for Quantum Design of Functional Materials, CAS Key Laboratory of Strongly-Coupled Quantum Matter Physics, and Department of Physics, University of Science and Technology of China, Hefei, Anhui 230026, China}
\author{Yafei Ren}
\affiliation{Department of Physics, University of Delaware, DE 19716, USA }
\author{Ying-Tao Zhang}
\email[Correspondence author:~~]{zhangyt@mail.hebtu.edu.cn}
\affiliation{College of Physics, Hebei Normal University, Shijiazhuang, Hebei 050024, China}
\author{Zhenhua Qiao}
\email[Correspondence author:~~]{qiao@ustc.edu.cn}
\affiliation{International Centre for Quantum Design of Functional Materials, CAS Key Laboratory of Strongly-Coupled Quantum Matter Physics, and Department of Physics, University of Science and Technology of China, Hefei, Anhui 230026, China}
\affiliation{Hefei National Laboratory, University of Science and Technology of China, Hefei 230088, China}
\author{Qian Niu}
\affiliation{International Centre for Quantum Design of Functional Materials, CAS Key Laboratory of Strongly-Coupled Quantum Matter Physics, and Department of Physics, University of Science and Technology of China, Hefei, Anhui 230026, China}
\date{\today}

\begin{abstract}
  We theoretically find that the second-order topological insulator, i.e., corner states, can be engineered by coupling two copies of two-dimensional $\mathbb{Z}_2$ topological insulators with opposite spin-helicities. As concrete examples, we utilize Kane-Mele models (i.e., graphene with intrinsic spin-orbit coupling) to realize the corner states by setting the respective graphenes to be $\mathbb{Z}_2$ topological insulators with opposite intrinsic spin-orbit couplings. To exhibit its universality, we generalize our findings to other representative $\mathbb{Z}_2$ topological insulators, e.g., the Bernevig-Hughes-Zhang model. An effective model is presented to reveal the physical origin of corner states. We further show that the corner states can also be designed in other topological systems, e.g., by coupling quantum anomalous Hall systems with opposite Chern numbers. Our work suggests that interlayer coupling can be treated as a simple and efficient strategy to drive lower-order topological insulators to the higher-order ones.
\end{abstract}

\maketitle
Topological insulators (TIs) represent a fascinating class of materials that exhibit insulating properties in their bulk while hosting topologically protected conducting states along their boundaries in two-dimensional (2D) systems or on their surfaces in three-dimensional (3D) systems~\cite{Haldane1988, Kane2005, Kane2005a, Bernevig2006, LiuFeng, Qi2011, Qiao2011, Qiao2010, Hasan2010, Bansil2016, Ren2016, Moore2010, Ando2013, Mong2010, Shen2024}. Rich topological phases have been identified in the past decades. Particularly, in the presence of time-reversal symmetry, $\mathbb{Z}_2$ TI~\cite{Olsen2019, Marrazzo2019, Choudhary2020, Slager2012, Kruthoff2017, Bradlyn2017, Po2017, Zhang2019, Vergniory2019, Tang2019} has been proposed and realized that is characterized by spin-helical edge states as represented by pioneering Kane-Mele model~\cite{Kane2005, Kane2005a} and Bernevig-Hughes-Zhang model~\cite{Bernevig2006} in 2D, which were later generalized to 3D~\cite{Fu2007, Roy2009}. In the absence of time-reversal symmetry, the quantum anomalous Hall effect (QAHE) with chiral-propagating edge modes have been widely studied in various different model systems and materials candidates~\cite{Liu2008, Qiao2010, Chang2023, Nagaosa2010, Weng2015, Xu2011, Liu2016, He2018, Chang2023, Mei2024, Yu2010, Wu2014, Fang2014, Qiao2014, Wang2014, Xu2015, Sun2019, Wang2013, Qi2016, Hogl2020, Devakul2022, Chang2013, Chang2015, Chang2015a, Deng2020, Serlin2020}.

Recent advancements have generalized the topological phases to higher order~\cite{Benalcazar2017, Benalcazar2017a, Li2020, Miert2018, Benalcazar2019, Schindler2019, Song2017, Schindler2018, Langbehn2017, Hsu2019, Schindler2018a, Peterson2018, Serra-Garcia2018, Xu2019, Franca2018, Kudo2019, Chen2020a, Yang2021}. In contrast to $\mathbb{Z}_2$ TI or QAHE, the topologically protected states in the higher-order TIs are localized at corners (0D states) in 2D systems or along hinges (1D states) in 3D systems~\cite{Benalcazar2017, Benalcazar2017a}. The emergence of higher-order TIs introduces a new dimension to the topological classification of materials, offering potential for novel electronic, photonic, and phononic applications~\cite{Xie2019, Zhu2021, Zhu2022, Huang2023}. In electronic systems, by case study, it was reported that 2D second-order TIs (SOTIs) can be realized in black phosphorene~\cite{Ezawa2018}, twisted bilayer graphene at certain angles~\cite{Park2019}, and graphyne~\cite{Liu2019, Lee2020, Sheng2019} where the corner states are spin degenerate due to the time-reversal symmetry. Besides, a systematical strategy to generate SOTIs is to break the time-reversal symmetry of the $\mathbb{Z}_2$ TIs by using in-plane Zeeman field~\cite{Ren2020} where the states on one corner is not degenerate, which was later applied to various materials systems~\cite{Huang2022, Zhuang2022, Han2022, Miao2022, Miao2023, Chen2020}.

\begin{figure}[tbp]
  \centering
  \includegraphics[width=8.5cm,angle=0]{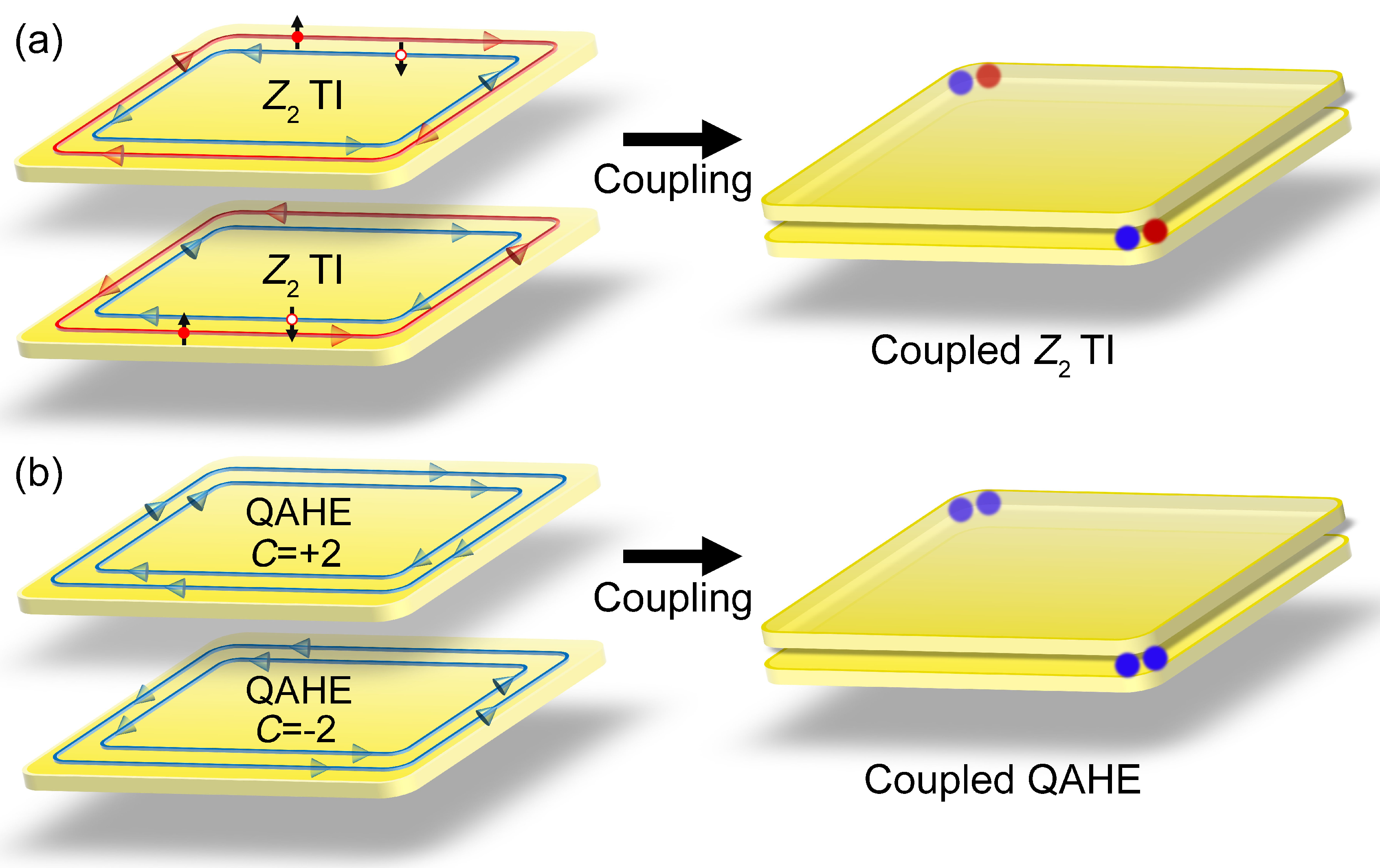}
  \caption{Schematic plot of coupled topological insulators. (a) Left: Two decoupled $\mathbb{Z}_2$ TI layers with opposite spin-helical edge states, i.e., spin-up edge modes in red respectively propagate clockwise and counterclockwise in the top and bottom layers; while spin-down edge modes in blue respectively propagate counterclockwise and clockwise in the top and bottom layers. Right: Coupling destroys all kinds of edge modes and gives rise to two Kramer pairs of corner states denoted by red and blue dots. (b) Left: Two decoupled QAHE layers with opposite Chern numbers, i.e., the edge modes propagate clockwise/counterclockwise in the top/bottom layer. Right: Coupling destroys all kinds of edge modes,and gives rise to two corner states denoted by blue dots.}
  \label{fig1}
\end{figure}

In this Letter, we demonstrate that, without breaking the time-reversal symmetry, the SOTI in 2D spinful systems can be engineered by simply coupling two first-order topological insulators, e.g., by coupling two copies of $\mathbb{Z}_2$ TI as illustrated in Fig.~\ref{fig1}(a). By using model study, we first demonstrate the SOTI by coupling two $\mathbb{Z}_2$ TI using the Kane-Mele model by identifying the in-gap corner states. Then, we generalize our results to the SOTI induced by coupling two QAHEs with opposite Chern numbers. Hereinbelow, we focus on the model systems on honeycomb lattice, where $\mathbb{Z}_2$ TI and QAHE can be realized by considering different ingredients~\cite{Kane2005, Qiao2010}. The generalization to other seminal model systems (Bernevig-Hughes-Zhang model) is provided in Ref.~[\onlinecite{LongPRB}].

\begin{figure}
  \centering
  \includegraphics[width=8.5cm,angle=0]{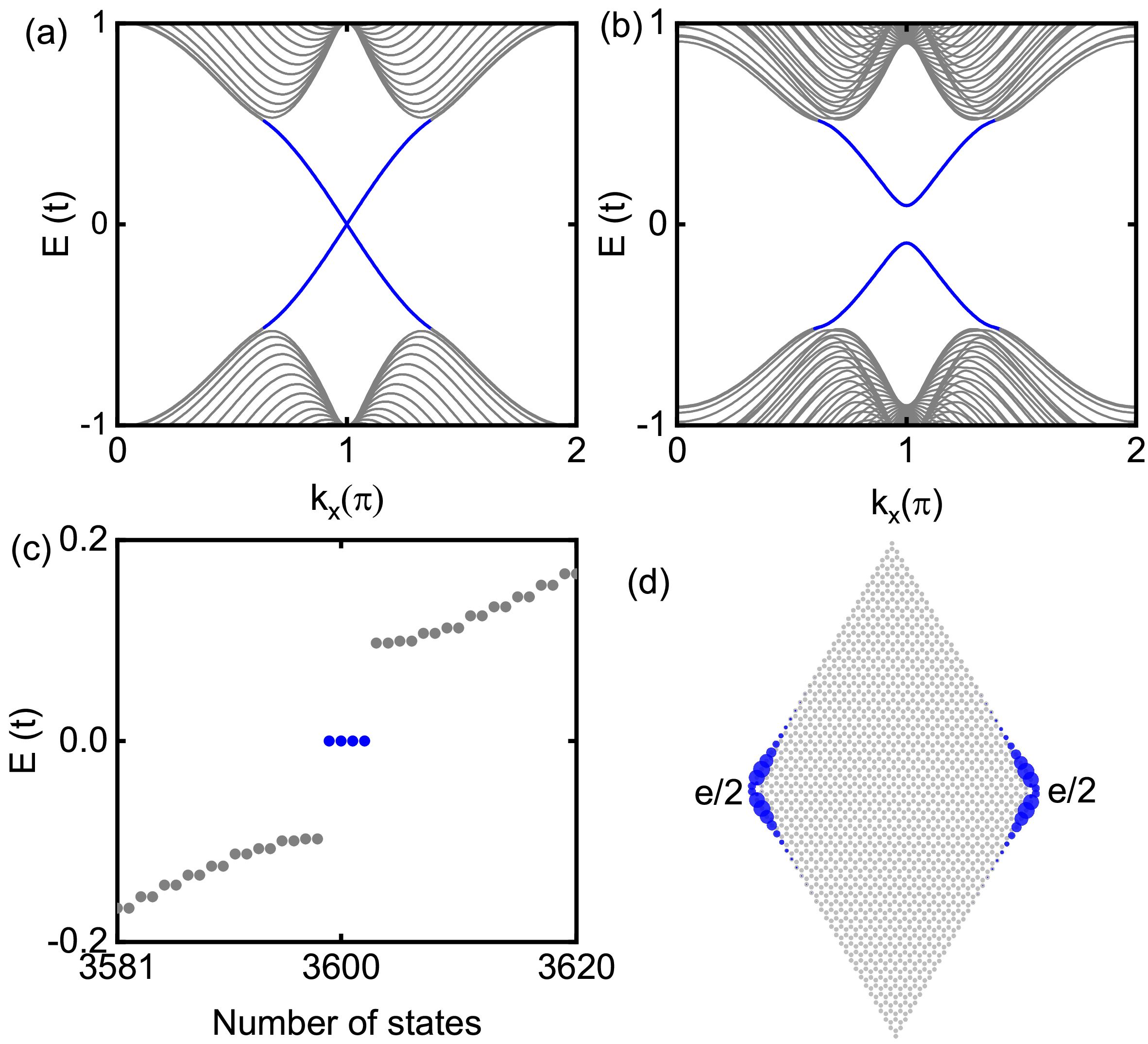}
  \caption{(a)-(b) Band structures of zigzag ribbons of decoupled and coupled $\mathbb{Z}_2$ TIs with coupling strengths being $\eta = 0.0$ (a) and $\eta = 0.2$ (b), respectively. Blue lines correspond to the spin-helical edge states. (c) Energy levels of diamond-shaped coupled graphene nanoflakes with the same parameters as those of panel (b). Blue dots correspond to in-gap corner states. (d) Probability distribution of the corner states. Other parameters are chosen to be $t^T_{\rm I} = -t^B_{\rm I}= 0.1$, $t_{\rm R}=0.0$, $\lambda=0.0$, $N_y = 60a$ for the ribbon width, and the nanoflake size $60a \times 60a$. 
  }
  \label{fig2}
\end{figure}

\textit{System Hamiltonian---.} The system Hamiltonian of the coupled two topological layers on the honeycomb lattice is shown below
\begin{align}
 H=
   \left(
   \begin{matrix}
         H_T &  \eta \\
         \eta ^*  &  H_B\\
   \end{matrix}
   \right),
   \label{eq1}
\end{align}
where $\eta$ measures the coupling strength between two separate layers, $H_T$ and $H_B$ are the modified Kane-Mele model Hamiltonians for top and bottom graphene layers, separately, which can be expressed as following:
\begin{eqnarray}
H_\gamma &=&-t\sum_{\langle ij \rangle}c^\dagger_{i}c_{j}+ i t_{\rm{R}} \sum_{{\langle ij \rangle} \alpha \beta} \hat{ \mathbf{e}}_z \cdot ({{\bf \sigma}_{\alpha\beta}} \times {\bf d}_{ij})c^{\dag}_{i\alpha}c_{j\beta} \nonumber
\\
&+&i t_{\rm{I}}^\gamma\sum_{\langle\langle ij \rangle\rangle}\nu_{ij}c^\dagger_{i}{s}_{z}c_{j}  + \lambda_\gamma \sum_{i\alpha}c^{\dagger}_{i\alpha} {\sigma_z} c_{i\alpha},
\label{EQ:SingleH}
\end{eqnarray}
where $\gamma$ is $T$ or $B$ representing top or bottom layer, $c^\dagger_{i}=(c^\dagger_{i\uparrow},c^\dagger_{i\downarrow})$ is the creation operator for an electron with spin up/down ($\uparrow$/$\downarrow$) at the $i$-th site. The first term is the nearest-neighbor hopping with an amplitude of $t$. The second term is the Rashba spin-orbit coupling with coupling strength $t_{\rm R}$. The third term is the intrinsic spin-orbit coupling involving next-nearest-neighbor hopping with $\nu_{ij}={\bm{d}_i \times \bm{d}_j}/{|\bm{d}_i \times \bm{d}_j|}$ where $\hat{\bm{d}}_{ij}$ is a unit vector pointing from site $j$ to $i$, with coupling strength $t_{\rm I}$. The last term corresponds to a uniform exchange field with a strength of $\lambda$. Without loss of generality, we only consider the AA-stacking case. Throughout this Letter, we measure the Fermi level, intrinsic and Rashba spin-orbit couplings, and exchange energy in the unit of $t$.

\begin{figure}
  \centering
  \includegraphics[width=8.5cm,angle=0]{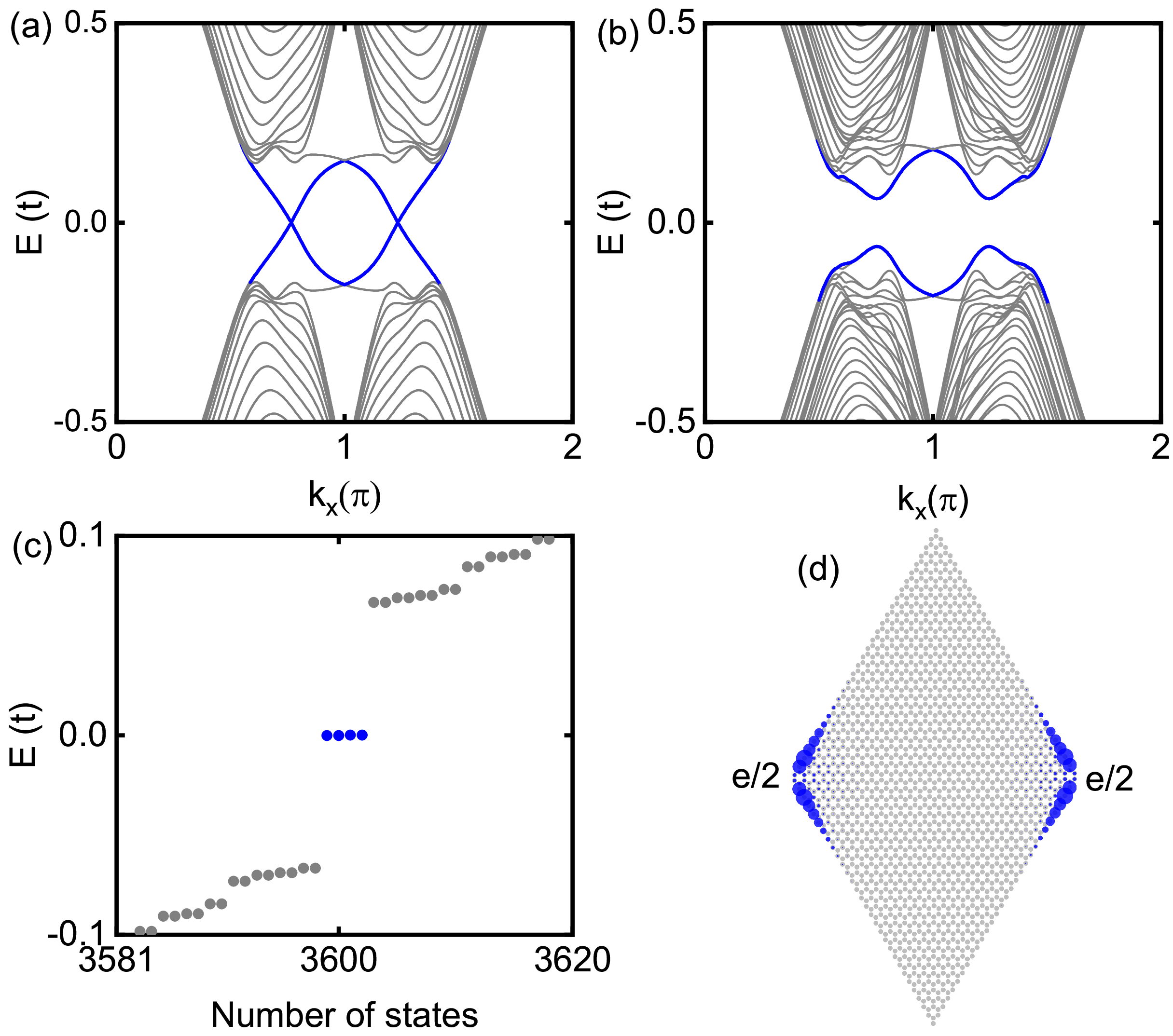}
  \caption{(a)-(b) Band structures of zigzag ribbons of decoupled and coupled QAHEs with coupling strengths being $\eta = 0.0$ (a) and $\eta = 0.1$ (b), respectively. Blue lines denote the chiral edge states.(c) Energy levels of diamond-shaped coupled graphene nanoflakes, with the same parameters as those of panel (b). Blue dots correspond to in-gap corner states. (d) Probability distribution of the corner states. Other parameters are chosen to be $t_{\rm I}=0.0$, $t_{\rm R}= 0.2$, $\lambda_{T}= -\lambda_{B}= 0.2$, $N_y = 60a$ for the ribbon width, and the nanoflake size $60a \times 60a$.
  }
  \label{fig3}
\end{figure}

\textit{Coupling $\mathbb{Z}_2$ TIs.---} Let us begin from the seminal Kane-Mele model, as depicted in the left panel of Fig.~\ref{fig1}(a), by setting $t_{\rm R}=\lambda_{T,B}=0$. In our consideration, we choose the zigzag graphene ribbon width to be $N_y = 60a$, with $a$ being the lattice constant. And we set the intrinsic spin-orbit couplings at top and bottom graphene layers to be opposite, i.e., $t^T_{\rm I}=-t^B_{\rm I}=0.1$. When the two layers are isolated (i.e., $\eta=0.0$), both exhibit the well-known spin-helical edge modes, i.e., edge modes with opposite spins counterpropagate along the same boundary [see Fig.~\ref{fig2}(a)]. The only difference is that, at top and bottom layers, the same spin-polarized edge mode propagates clockwise and counterclockwise, respectively [see the left panel of Fig.~\ref{fig1}(a)].

The stacking of two $\mathbb{Z}_2$ TIs is $\mathbb{Z}_2$ topologically trivial as the gapless edge states from the same spin-sector can couple and form an energy gap. As shown in Fig.~\ref{fig2}(b) where the edge states highlighted in blue become gapped when the interlayer coupling is turned on ($\eta = 0.2$). A natural and general consequence is that ``+" TI phase plus ``-" TI phase gives nothing, i.e. a trivial insulator. Surprisingly, we show that although the first-order topological phase vanishes, the second-order topological phase arises.

To explore the second-order topology, an efficient approach is to explore the energy spectra of a fixed nanoflake, e.g., a diamond-shaped nanoflake ($60a \times 60a$) for graphene systems. As displayed in Fig.~\ref{fig2}(c), there arise four zero-energy in-gap states (highlighted in blue dots) at the Fermi level. By analyzing the local density of states of the zero-energy states, one can see that these states are equally localized at the left and right corners, {with each corner hosting 1/2 electron charge}. This directly indicates the formation of the SOTIs in 2D systems. Some other various representative nanoflakes are considered in Ref.~[\onlinecite{LongPRB}]. We also successfully generalize these findings to other seminal $\mathbb{Z}_2$ TIs, e.g., the Bernevig-Hughes-Zhang (BHZ) model, as discussed in Ref.~[\onlinecite{LongPRB}].

\textit{Physical origin of corner states.---} To understand the physical origin, let us first recall the effective Hamiltonian of Kane-Mele model with in-plane magnetic field~\cite{Ren2020} on the basis of \{$\psi_{\uparrow}, \psi_{\downarrow}$\}, which is written as~\cite{Shen2017}:
\begin{eqnarray}
h_{0}=\left[ \begin{matrix}
 \textbf{d}(\textbf{k}) \cdot \sigma &  B_x \\
 B_x  &  -\textbf{d}(-\textbf{k}) \cdot \sigma \\
\end{matrix} \right],
 \label{eq-eff1}
\end{eqnarray}
where ${\textbf{d}(\textbf{k})} \cdot \sigma$ and $-\textbf{d}(-\textbf{k}) \cdot \sigma$ represent the spin-up and spin-down sectors, respectively. And, $\mathcal{T} \textbf{d(k)} \cdot \sigma \mathcal{T}^{-1} = -\textbf{d}(-\textbf{k}) \cdot \sigma$, where the time-reversal operator $\mathcal{T} = i s_y \mathcal{K}$. $B_x$ is the in-plane Zeeman field, which couples the spin-up and spin-down edge modes propagating along opposite directions, therefore induces the edge gap to harbour the corner states.

In the present work, the same spin-polarized edge states propagate in opposite directions at different layers, with the effective Hamiltonian on the basis of \{$\psi_{T\uparrow}, \psi_{T\downarrow}, \psi_{B\uparrow}, \psi_{B\downarrow}$\} being written as:
\begin{eqnarray}
 h_{1}  =\left[ \begin{matrix}
\textbf{d}_T(\textbf{k}) \cdot \sigma &  0 &  \eta  &  0 \\
 0  &  -\textbf{d}_T(-\textbf{k}) \cdot \sigma &  0 &  \eta\\
 \eta  &  0 &  \textbf{d}_B(-\textbf{k}) \cdot \sigma  &  0 \\
 0 &  \eta &  0  &  -\textbf{d}_B(\textbf{k}) \cdot \sigma\\
\end{matrix} \right]\nonumber,
 \label{eq-eff2}
\end{eqnarray}
where $T$ and $B$ represent top and bottom Kane-Mele system, respectively. Since the spin is a good quantum number in Kane-Mele model, one can consider the single spin sectors separately, and the Hamiltonian on the basis of \{$\psi_{T\uparrow}, \psi_{B\uparrow}$\} can be written as:
\begin{eqnarray}
   H=\left[
   \begin{matrix}
 \textbf{d}_T\textbf{(k)} \cdot \sigma &  \eta \\
 \eta  &  \textbf{d}_B(-\textbf{k}) \cdot \sigma \\
\end{matrix} \right].
 \label{eq-eff3}
\end{eqnarray}
Since $t^T_{\rm I}=-t^B_{\rm I}$, the Eqs.~(\ref{eq-eff1}) and ~(\ref{eq-eff3}) share the same form, except that the in-plane Zeeman field in Eq.~(\ref{eq-eff1}) is replaced by the interlayer coupling in Eq.~(\ref{eq-eff3}).
In Eq.~(\ref{eq-eff3}) the interlayer coupling couples the same spin-polarized but counterpropagating edge modes at different layers to open edge gaps to form the corner states. It is highly noteworthy that in the coupled TIs system the time-reversal symmetry is always preserved, i.e., the in-plane magnetization is not required.

\textit{Coupling QAHEs.---} Next, we show that the above strategy can also apply to QAHEs. In the following, we demonstrate the presence of SOTI by coupling two QAHE layers with opposite Chern numbers by setting $t_{\rm I}=0.0$, $t_{\rm R}= 0.2$, and $\lambda_{T}= -\lambda_{B}= 0.2$. When the two layers are isolated (i.e., $\eta=0.0$), the top and bottom graphene layer give rise to QAHEs with opposite Chern numbers of $\mathcal{C}=\pm2$. As plotted in Fig.~\ref{fig3}(a), the band structure of the zigzag ribbons exhibits doubly degenerate and chirally-propagating gapless edge modes as shown in blue (clockwise and counterclockwise respectively at top and bottom layers as depicted in Fig.~\ref{fig1}(b)). When the coupling is turned on with $\eta = 0.1$, the gapless edge mode becomes gapped as shown in Fig.~\ref{fig3}(b). To confirm the presence of SOTI, we plot the energy spectra of the diamond-shaped nanoflakes as displayed in Fig.~\ref{fig3}(c). Both the emergence of four zero-energy states and their wavefunction distribution at the corners [see Fig.~\ref{fig3}(d)] together strongly indicate the existence of SOTI. Similarly, this finding can also be generalized to other representative QAHE systems, e.g., BHZ model with Zeeman field~\cite{Liu2008, LongPRB}, and the Dirac model~[\onlinecite{LongPRB}].

\begin{figure}
  \centering
  \includegraphics[width=8.5cm,angle=0]{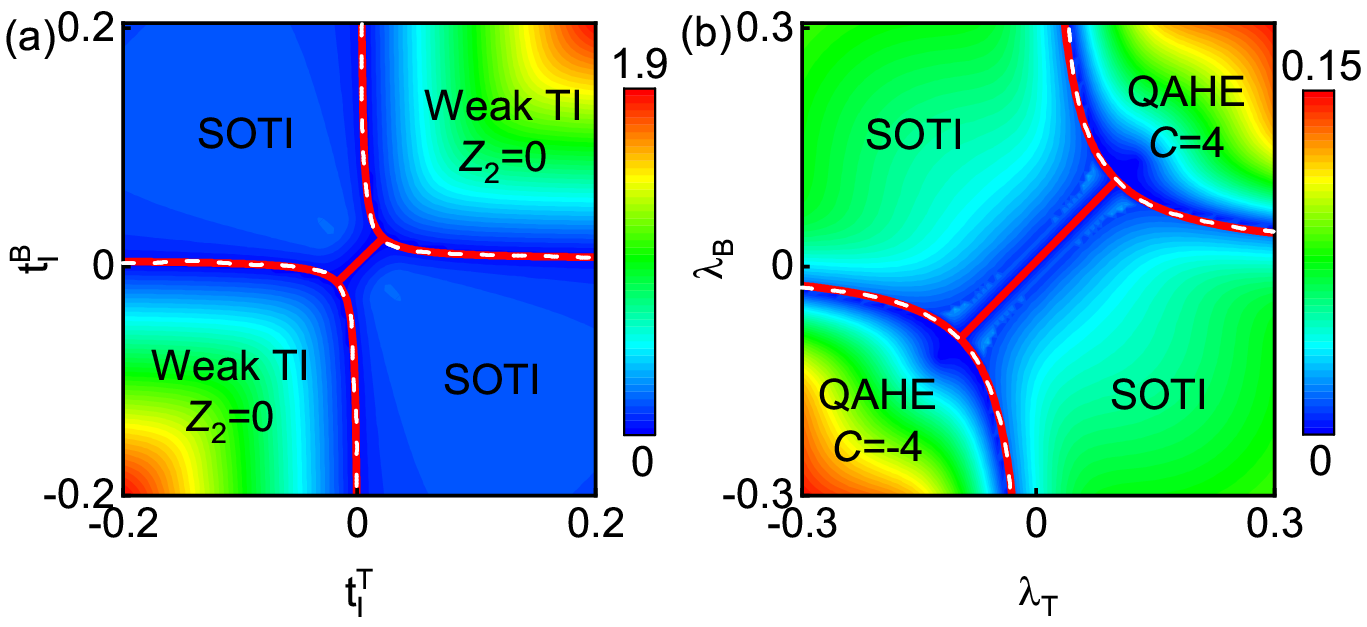}
  \caption{ (a) Phase diagram as functions of the intrinsic spin-orbit couplings $t^T_{\rm I}$ and $t^B_{\rm I}$ for the coupled graphene layers. Other parameters are set to be $t_{\rm R}=0.0$, $\lambda=0.0$, and $\eta=0.2$.
  (b) Phase diagram as functions of the exchange fields $\lambda_{T}$ and $\lambda_{B}$ for the coupled graphene layers.
  Other parameters are set to be $t_{\rm R}=0.2$, $t_{\rm I}=0.0$, and $\eta=0.1$. Dotted lines are the phase boundaries.
  Here, the bulk gap is used for the TI, weak TI and QAHE regions, while the edge gap is used for the SOTI region.}
  \label{fig4}
\end{figure}

\textit{Topological Phase diagrams.---} In above discussions, we have chosen the specific case where the determining parameters are exactly opposite, i.e. $t^{T}_{\rm I}=-t^{B}_{\rm I}$  for coupled $\mathbb{Z}_2$ TI or $\lambda_{T}=-\lambda_{B}$ for coupled QAHE systems.
Here we systematically explore the influence of the relative signs on the topological phases. Figure~\ref{fig4}(a) shows the topological phase diagram in the parameter spaces of ($t^T_{\rm I}$, $t^B_{\rm I}$). One can see that there are four regions, which respectively belong to SOTI phases and weak TI phases. One can see that as long as the signs of $t^T_{\rm I}$ and $t^B_{\rm I}$ are different, any coupling can lead to the formation of SOTI (see second and fourth quadrants). But, when the signs of $t^T_{\rm I}$ and $t^B_{\rm I}$ are identical, the coupling can result in either weak TI or SOTI. A similar phase diagram as plotted in Fig.~\ref{fig4}(b) is obtained for the coupled QAHE systems. For example, when $\lambda_{T}$ and $\lambda_{B}$ have different signs, the system is driven into SOTI; but when they have the identical sign, it can induce either SOTI or QAHE with $\mathcal{C}=\pm 4$.

To clearly understand and determine the phase boundaries, we employ the low-energy continuum model expanded at valley K/K':
\begin{align}
 H_{\rm eff}=
   \left[
   \begin{matrix}
          H^T_{\rm eff} &  \eta\sigma_0 s_0 \\
         \eta\sigma_0 s_0  &  H^B_{\rm eff}\\
   \end{matrix}
   \right],
   \label{eq3}
\end{align}
where
\begin{eqnarray}
H^{T,B}_{\rm eff}&=&3t/2 (\sigma_x k_x +\sigma_y k_y)\mathbf{1}_s +3t_{\rm R}/2 (\sigma_x s_y-\sigma_y s_x) \nonumber \\
&+&3\sqrt{3}t^{T,B}_{\rm I}\sigma_z s_z+\lambda_{T,B}s_z \mathbf{1}_\sigma.
\end{eqnarray}
For the weak TI and SOTI phase boundaries, we set $t_{\rm R}=0$, $\lambda_{T,B}=0$, and impose the bulk gap closing condition of $\varepsilon=0$. One can obtain the topological phase transition boundary satisfying
\begin{eqnarray}
t^T_{\rm I}t^B_{\rm I}=\eta^2/27,
\end{eqnarray}
which shows perfect consistence with the direct band structure calculation. Similarly, for the QAHE and SOTI phase boundaries, by setting $t_{\rm I}=0$, the topological phase transition boundary satisfies
\begin{eqnarray}
\lambda_T \lambda_B=\eta^2,
\end{eqnarray}
which also agrees well with the direct band structure calculation from tight-binding model.

\textit{Conclusions.---}
We have shown that the SOTIs in spinful systems can be engineered without breaking the time-reversal symmetry, by simply introducing an interlayer coupling in two $\mathbb{Z}_2$ TIs with opposite spin helicities within the framework of Kane-Mele model. A minimal model was presented to help understand the formation of SOTIs, i.e., the interaction between the counterpropagating edge modes with the same spin at different layers opens the edge gap. Inspired by these findings, we have also found that coupling two QAHEs with opposite Chern numbers can also lead to the formation of SOTIs, by using the graphene model with Rashba effect. The generalization of these findings to the BHZ model suggests the universal characteristic of our proposed strategy of engineering SOTIs by coupling any kinds of 2D topological systems.

\textit{Acknowledgements.---}
This work was financially supported by the National Natural Science Foundation of China (Grants No. 12074097, No. 11974327, and No. 12004369), Natural Science Foundation of Hebei Province (Grant No. A2020205013), Anhui Initiative in Quantum Information Technologies (Grant No. AHY170000), and Innovation Program for Quantum Science and Technology (Grant No. 2021ZD0302800). The supercomputing service of USTC is gratefully acknowledged.

* These authors contribute equally to this work.


\begin{thebibliography}{99}
\bibitem{Haldane1988} F. D. M. Haldane, Phys. Rev. Lett. \textbf{61}, 2015 (1988).
\bibitem{Kane2005} C. L. Kane and E. J. Mele, Phys. Rev. Lett. \textbf{95}, 146802 (2005).
\bibitem{Kane2005a} C. L. Kane and E. J. Mele, Phys. Rev. Lett. \textbf{95}, 226801 (2005).
\bibitem{Bernevig2006} B. A. Bernevig, T. L. Hughes, and S.-C. Zhang, Science \textbf{314}, 1757 (2006).
\bibitem{Moore2010} J. E. Moore, Nature (London) \textbf{464}, 194 (2010).
\bibitem{Ando2013} Y. Ando, J. Phys. Soc. Jpn. \textbf{82}, 102001 (2013).
\bibitem{Mong2010} R. S. K. Mong, A. M. Essin, and J. E. Moore, Phys. Rev. B \textbf{81}, 245209 (2010).
\bibitem{LiuFeng} F. Liu, Coshare Science \textbf{01}, v3, 1-62 (2023). DOI: https://doi.org/10.61109/cs.202310.115.
\bibitem{Qi2011} X.-L. Qi and S.-C. Zhang, Rev. Mod. Phys. \textbf{83}, 1057 (2011).
\bibitem{Qiao2011} Z. Qiao, W.-K. Tse, H. Jiang, Y. G Yao, and Q. Niu, Phys. Rev. Lett. \textbf{107}, 256801 (2011).
\bibitem{Qiao2010} Z. Qiao, S. A. Yang, W. Feng, W. K. Tse, J. Ding, Y. Yao, J. Wang, and Q. Niu, Phys. Rev. B \textbf{82}, 161414(R) (2010).
\bibitem{Hasan2010} M. Z. Hasan and C. L. Kane, Rev. Mod. Phys. \textbf{82}, 3045 (2010).
\bibitem{Bansil2016} A. Bansil, H. Lin, and T. Das, Rev. Mod. Phys. \textbf{88}, 021004 (2016).
\bibitem{Ren2016} Y. Ren, Z. Qiao, and Q. Niu, Rep. Prog. Phys. \textbf{79}, 066501 (2016).
\bibitem{Shen2024} S. Q. Shen, Coshare Science \textbf{02}, v1, 1-42 (2024). DOI: https://doi.org/10.61109/cs.202402.128.
\bibitem{Olsen2019} T. Olsen, E. Andersen, T. Okugawa, D. Torelli, T. Deilmann, and K. S. Thygesen, Phys. Rev. Mater. \textbf{3}, 024005 (2019).
\bibitem{Marrazzo2019} A. Marrazzo, M. Gibertini, D. Campi, N. Mounet, and N. Marzari, Nano Lett. \textbf{19}, 8431 (2019).
\bibitem{Choudhary2020} K. Choudhary, K. F. Garrity, J. Jiang, R. Pachter, and F. Tavazza, npj Computational Mater. \textbf{6}, 49 (2020).
\bibitem{Slager2012} R.-J. Slager, A. Mesaros, V. Juri{\v{c}}i{\'{c}}, and J. Zaanen, Nat. Phys. \textbf{9}, 98 (2012).
\bibitem{Kruthoff2017} J. Kruthoff, J. de Boer, J. van Wezel, C. L. Kane, and R.-J. Slager, Phys. Rev. X \textbf{7}, 041069 (2017).
\bibitem{Bradlyn2017} B. Bradlyn, L. Elcoro, J. Cano, M. G. Vergniory, Z. Wang, C. Felser, M. I. Aroyo, and B. A. Bernevig, Nature \textbf{547}, 298 (2017).
\bibitem{Po2017} H. C. Po, A. Vishwanath, and H. Watanabe, Nat. Commun. \textbf{8}, 50 (2017).
\bibitem{Zhang2019} T. Zhang, Y. Jiang, Z. Song, H. Huang, Y. He, Z. Fang, H. Weng, and C. Fang, Nature \textbf{566}, 475 (2019).
\bibitem{Vergniory2019} M. G. Vergniory, L. Elcoro, C. Felser, N. Regnault, B. A. Bernevig, and Z. Wang, Nature \textbf{566}, 480 (2019).
\bibitem{Tang2019} F. Tang, H. C. Po, A. Vishwanath, and X. Wan, Nature \textbf{566}, 486 (2019).

\bibitem{Fu2007} L. Fu, C. L. Kane, and E. J. Mele, Phys. Rev. Lett. \textbf{98}, 106803 (2007).
\bibitem{Roy2009} R. Roy, Phys. Rev. B \textbf{79}, 195322 (2009).

\bibitem{Liu2008} C.-X. Liu, X.-L. Qi, X. Dai, Z. Fang, and S.-C. Zhang, Phys. Rev. Lett. \textbf{101}, 146802 (2008).
\bibitem{Nagaosa2010} N. Nagaosa, J. Sinova, S. Onoda, A. H. MacDonald, and N. P. Ong, Rev. Mod. Phys. \textbf{82}, 1539 (2010).
\bibitem{Weng2015} H. Weng, R. Yu, X. Hu, X. Dai, and Z. Fang, Adv. Phys. \textbf{64}, 227 (2015).
\bibitem{Xu2011} G. Xu, H. Weng, Z. Wang, X. Dai, and Z. Fang, Phys. Rev. Lett. \textbf{107}, 186806 (2011).
\bibitem{Liu2016} C.-X. Liu, S.-C. Zhang, and X.-L. Qi, Annu. Rev. Condens. Matter Phys. \textbf{7}, 301 (2016).
\bibitem{He2018} K. He, Y. Wang, and Q.-K. Xue, Annu. Rev. Condens. Matter Phys. \textbf{9}, 329 (2018).
\bibitem{Chang2023} C.-Z. Chang, C.-X. Liu, and A. H. MacDonald, Rev. Mod. Phys. \textbf{95}, 011002 (2023).
\bibitem{Mei2024} R. Mei, Y.-F. Zhao, C. Wang, Y. Ren, D. Xiao, C.-Z. Chang, and C.-X. Liu, Phys. Rev. Lett. \textbf{132}, 066604 (2024).

\bibitem{Yu2010} R. Yu, W. Zhang, H.-J. Zhang, S.-C. Zhang, X. Dai, and Z. Fang, Science \textbf{329}, 61 (2010).
\bibitem{Wu2014} S.-C. Wu, G. Shan, and B. Yan, Phys. Rev. Lett. \textbf{113}, 256401 (2014).
\bibitem{Fang2014} C. Fang, M. J. Gilbert, and B. A. Bernevig, Phys. Rev. Lett. \textbf{112}, 046801 (2014).
\bibitem{Qiao2014} Z. Qiao, W. Ren, H. Chen, L. Bellaiche, Z. Zhang, A. H. MacDonald, and Q. Niu, Phys. Rev. Lett. \textbf{112}, 116404 (2014).
\bibitem{Wang2014} Q.-Z. Wang, X. Liu, H.-J. Zhang, N. Samarth, S.-C. Zhang, and C.-X. Liu, Phys. Rev. Lett. \textbf{113}, 147201 (2014).
\bibitem{Xu2015} G. Xu, B. Lian, and S.-C. Zhang, Phys. Rev. Lett. \textbf{115}, 186802 (2015).
\bibitem{Sun2019} H. Sun, B. Xia, Z. Chen, Y. Zhang, P. Liu, Q. Yao, H. Tang, Y. Zhao, H. Xu, and Q. Liu, Phys. Rev. Lett. \textbf{123}, 096401 (2019).
\bibitem{Wang2013} Z. F. Wang, Z. Liu, and F. Liu, Phys. Rev. Lett. \textbf{110}, 196801 (2013).
\bibitem{Qi2016} S. Qi, Z. Qiao, X. Deng, E. D. Cubuk, H. Chen, W. Zhu, E. Kaxiras, S. B. Zhang, X. Xu, and Z. Zhang, Phys. Rev. Lett. \textbf{117}, 056804 (2016).
\bibitem{Hogl2020} P. H$\rm{\ddot{o}}$gl, T. Frank, K. Zollner, D. Kochan, M. Gmitra, and J. Fabian, Phys. Rev. Lett. \textbf{124}, 136403 (2020).
\bibitem{Devakul2022} T. Devakul and L. Fu, Phys. Rev. X \textbf{12}, 021031 (2022)
\bibitem{Chang2013} C.-Z. Chang, J. Zhang, X. Feng, J. Shen, Z. Zhang, M. Guo, K. Li, Y. Ou, P. Wei, L.-L. Wang, Z.-Q. Ji, Y. Feng, S. Ji, X. Chen, J. Jia, X. Dai, Z. Fang, S.-C. Zhang, K. He, Y. Wang et al., Science \textbf{340}, 167 (2013)
\bibitem{Chang2015} C.-Z. Chang, W. Zhao, D. Y. Kim, P. Wei, J. K. Jain, C. Liu, M. H. W. Chan, and J. S. Moodera, Phys. Rev. Lett. \textbf{115}, 057206 (2015).
\bibitem{Chang2015a} C.-Z. Chang, W. Zhao, D. Y. Kim, H. Zhang, B. A. Assaf, D. Heiman, S.-C. Zhang, C. Liu, M. H. W. Chan, and J. S. Moodera, Nature Mater. \textbf{14}, 473 (2015).
\bibitem{Deng2020} Y. Deng, Y. Yu, M. Z. Shi, Z. Guo, Z. Xu, J. Wang, X. H. Chen, and Y. Zhang, Science \textbf{367}, 895 (2020).
\bibitem{Serlin2020} M. Serlin, C. L. Tschirhart, H. Polshyn, Y. Zhang, J. Zhu, K. Watanabe, T. Taniguchi, L. Balents, and A. F. Young, Science \textbf{367}, 900 (2020).
\bibitem{Benalcazar2017} W. A. Benalcazar, B. A. Bernevig, and T. L. Hughes, Science \textbf{357}, 61 (2017).
\bibitem{Benalcazar2017a} W. A. Benalcazar, B. A. Bernevig, and T. L. Hughes, Phys. Rev. B \textbf{96}, 245115 (2017).
\bibitem{Li2020} T. Li, P. Zhu, W. A. Benalcazar, and T. L. Hughes, Phys. Rev. B \textbf{101}, 115115 (2020).
\bibitem{Miert2018} G. van Miert and C. Ortix, Phys. Rev. B \textbf{98}, 081110(R) (2018).
\bibitem{Benalcazar2019} W. A. Benalcazar, T. Li, and T. L. Hughes, Phys. Rev. B \textbf{99}, 245151 (2019).
\bibitem{Schindler2019} F. Schindler, M. Brzezinska, W. A. Benalcazar, M. Iraola, A. Bouhon, S. S. Tsirkin, M. G. Vergniory, Phys. Rev. Res. \textbf{1}, 033074 (2019).
\bibitem{Song2017} Z. Song, Z. Fang, and C. Fang, Phys. Rev. Lett. \textbf{119}, 246402 (2017).
\bibitem{Schindler2018} F. Schindler, A. M. Cook, M. G. Vergniory, Z. Wang, S. S. P. Parkin, B. A. Bernevig, and T. Neupert, Sci. Adv. \textbf{4}, eaat0346 (2018).
\bibitem{Langbehn2017} J. Langbehn, Y. Peng, L. Trifunovic, F. von Oppen, and P. W. Brouwer, Phys. Rev. Lett. \textbf{119}, 246401 (2017).
\bibitem{Hsu2019} C.-H. Hsu, X. Zhou, T.-R. Chang, Q. Ma, N. Gedik, A. Bansil, S.-Y. Xu, H. Lin, and L. Fu, Proc. Natl. Acad. Sci. U.S.A. \textbf{116}, 13255 (2019).
\bibitem{Schindler2018a} F. Schindler, Z. Wang, M. G. Vergniory, A. M. Cook, A. Murani, S. Sengupta, A. Y. Kasumov, R. Deblock, S. Jeon, I. Drozdov, H. Bouchiat, S. Guron, A. Yazdani, B. A. Bernevig, and T. Neupert, Nat. Phys. \textbf{14}, 918 (2018).
\bibitem{Peterson2018} C. W. Peterson, W. A. Benalcazar, T. L. Hughes, and G. Bahl, Nature \textbf{555}, 346 (2018).
\bibitem{Serra-Garcia2018} M. Serra-Garcia, V. Peri, R. Ssstrunk, O. R. Bilal, T. Larsen, L. G. Villanueva, and S. D. Huber, Nature \textbf{555}, 342 (2018).
\bibitem{Xu2019} Y. Xu, Z. Song, Z. Wang, H. Weng, and X. Dai, Phys. Rev. Lett. \textbf{122}, 256402 (2019).
\bibitem{Franca2018} S. Franca, J. van den Brink, and I. C. Fulga, Phys. Rev. B \textbf{98}, 201114(R) (2018).
\bibitem{Kudo2019} K. Kudo, T. Yoshida, and Y. Hatsugai, Phys. Rev. Lett. \textbf{123}, 196402 (2019).
\bibitem{Chen2020a} R. Chen, C.-Z. Chen, J.-H. Gao, B. Zhou, and D.-H. Xu, Phys. Rev. Lett. \textbf{124}, 036803 (2020).
\bibitem{Yang2021} Y.-B. Yang, K. Li, L.-M. Duan, and Y. Xu, Phys. Rev. B \textbf{103}, 085408 (2021).
\bibitem{Xie2019} B. Xie, G. Su, H. Wang, H. Su, X. Shen, P. Zhan, M. Lu, Z. Wang, and Y. Chen, Phys. Rev. Lett. \textbf{122}, 233903 (2019).
\bibitem{Zhu2021} B. Zhu, Q. Wang, Y. Zeng, Q. J. Wang, and Y. D. Chong, Phys. Rev. B \textbf{104}, L140306 (2021).
\bibitem{Zhu2022} J. Zhu, W. Wu, J. Zhao, C. Chen, Q. Wang, X.-L. Sheng, L. Zhang, Y. X. Zhao, and S. A. Yang, Phys. Rev. B \textbf{105}, 085123 (2022).
\bibitem{Huang2023} F. Huang, P. Zhou, W. Li, S. He, R. Tan, Z. Ma, and L. Z. Sun, Phys. Rev. B \textbf{107}, 134104 (2023).
\bibitem{Ezawa2018} M. Ezawa, Phys. Rev. B \textbf{98}, 045125 (2018).
\bibitem{Park2019} M. J. Park, Y. Kim, G. Y. Cho, and S. Lee, Phys. Rev. Lett. \textbf{123}, 216803 (2019).
\bibitem{Liu2019} B. Liu, G. Zhao, Z. Liu, and Z. F. Wang, Nano Lett. \textbf{19}, 6492 (2019).
\bibitem{Lee2020} E. Lee, R. Kim, J. Ahn, and B.-J. Yang, npj Quantum Mater. \textbf{5}, 1 (2020).
\bibitem{Sheng2019} X.-L. Sheng, C. Chen, H. Liu, Z. Chen, Z.-M. Yu, Y. Zhao, and S. A. Yang, Phys. Rev. Lett. \textbf{123}, 256402 (2019).
\bibitem{Ren2020} Y. Ren, Z. Qiao, and Q. Niu, Phys. Rev. Lett. \textbf{124}, 166804 (2020).
\bibitem{Huang2022} X. Huang, J. Lu, Z. Yan, M. Yan, W. Deng, G. Chen, and Z. Liu, Sci. Bull. \textbf{67}, 488 (2022).
\bibitem{Zhuang2022} Z.-Y. Zhuang and Z. Yan, Front. Phys. \textbf{10}, 866347 (2022)
\bibitem{Han2022} B. Han, J. Zeng, and Z. Qiao, Chinese Phys. Lett. \textbf{39}, 017302 (2022).
\bibitem{Miao2022} C.-M. Miao, Q.-F. Sun, and Y.-T. Zhang, Phys. Rev. B \textbf{106}, 165422 (2022).
\bibitem{Miao2023} C.-M. Miao, Y.-H. Wan, Q.-F. Sun, and Y.-T. Zhang, Phys. Rev. B \textbf{108}, 075401 (2023).
\bibitem{Chen2020} C. Chen, Z. Song, J.-Z. Zhao, Z. Chen, Z.-M. Yu, X.-L. Sheng, and S. A. Yang, Phys. Rev. Lett. \textbf{125}, 056402 (2020).
\bibitem{LongPRB} L. Liu, J. An, Y. F. Ren, Y.-T. Zhang, and Z. Qiao, in preparation [2024].
\bibitem{Shen2017} S.-Q Shen, \textit{Topological Insulators: Dirac Equation in Condensed Matter}, 2nd ed. (Springer-Verlag, Berlin, 2017).

\end{thebibliography}
\end{document}